\documentclass{emulateapj}
\usepackage{apjfonts}

\slugcomment{To appear in {\it The Astrophysical Journal}.}
\shorttitle{SCATTERED X-RAYS IN OBSCURED AGNs}
\shortauthors{NOGUCHI ET AL.}

\begin{document}
\title{Scattered X-rays in Obscured Active Galactic Nuclei and their Implications for Geometrical Structure and Evolution}
\author{Kazuhisa Noguchi\altaffilmark{1}, Yuichi Terashima\altaffilmark{1},  Yukiko Ishino\altaffilmark{2}, Yasuhiro Hashimoto\altaffilmark{3}, Michael Koss\altaffilmark{4, 5},  Yoshihiro Ueda\altaffilmark{2},  and Hisamitsu Awaki\altaffilmark{1}
}
\altaffiltext{1}{Department of Physics, Ehime University, Matsuyama, Ehime 790-8577, Japan.}
\altaffiltext{2}{Department of Astronomy, Kyoto University, Kyoto 606-8502, Japan.}
\altaffiltext{3}{Department of Earth Sciences, National Taiwan Normal University, No.88, Sec. 4, Tingzhou Rd., Wenshan District, Taipei 11677, Taiwan}
\altaffiltext{4}{Astrophysics Science Division, X-Ray Astrophysical Laboratory, NASA Goddard Space Flight Center, Greenbelt, MD 20771, USA}
\altaffiltext{5}{Department of Astronomy, University of Maryland, College Park, MD 20742, USA}

\begin{abstract}

We construct a new sample of 32 obscured active galactic nuclei (AGNs)
selected from the Second {\it XMM-Newton} Serendipitous Source
Catalogue to investigate their multiwavelength properties in relation
to the ``scattering fraction", the ratio of the soft X-ray flux to the
absorption-corrected direct emission. The sample covers a broad range
of the scattering fraction ($\sim$0.1\%$-$10\%). A quarter of the 32 AGNs
have a very low scattering fraction ($\leq0.5$\%), which suggests that
they are buried in a geometrically thick torus with a very small
opening angle. We investigate correlations between the scattering
fraction and multiwavelength properties. We find that AGNs with a small
scattering fraction tend to have low [\ion{O}{3}]$\lambda$5007/X-ray
luminosity ratios. This result agrees with the expectation that the
extent of the narrow-line region is small because of the small opening
angle of the torus. There is no significant correlation between
scattering fraction and far-infrared luminosity. This implies that a
scale height of the torus is not primarily determined by starburst
activity. We also compare scattering fraction with black hole mass or
Eddington ratio and find a weak anti-correlation between the Eddington
ratio and scattering fraction. This implies that more rapidly growing 
supermassive black holes tend to have thicker tori. 
\end{abstract}

\section{Introduction}
Understanding the cosmological evolution of active galactic nuclei
(AGNs) is an important issue in modern astrophysics and is closely 
related to the evolution of galaxies. One piece of evidence for ``co-evolution" of 
AGNs and galaxies is the strong correlation between a central black hole 
and its host galaxy (e.g., Magorrian et al. 1998; Marconi \& Hunt 2003).
In an early stage of galaxy evolution, the nuclear region may be hidden
behind rich gas, which leads to active star formation. Indeed
recent models predict that the central black holes in all galaxies
experience a heavily obscured phase (Hopkins et al. 2006,
2008). Therefore, investigating obscured AGNs is of significant
interest to study the co-evolution of black holes and galaxies.
Moreover, previous observations have shown that obscured AGNs are 
a major population of the AGN.
Finally, obscured AGN 
are predicted to be the main contributors to the unexplained 
cosmic X-ray background (CXB) in the hard X-rays 
(Comastri et al. 1995; Ueda et al. 2003; Gilli et al. 2007).
Thus, obscured AGN population is a key class
to understand the overall AGN population, AGN evolution, and
relationship between black holes and their hosts.

According to a unified model of the AGN, obscuring matter referred to as a
``torus'' is surrounding a supermassive black hole and gas
photoionized by the AGN is created in the opening part of the torus
(Antonucci 1993). If we observe an AGN from the torus side, absorbed
direct emission from the nucleus and emission scattered by the
photoionized gas are observed in the X-ray spectra.  A scattering
fraction is calculated as the fraction of the scattered emission with
respect to the direct emission and reflects the solid angle of the
opening part of the torus and/or an amount of gas responsible for
scattering. {\it Suzaku's} follow-up observations of {\it Swift}
BAT-detected AGNs found a new type of AGN with a very small scattering
fraction ($<$ 0.5\%; Ueda et al. 2007; Eguchi et al. 2009; Winter et
al. 2009). Assuming that the amount of scattering medium does not
differ much from object to object, they would be buried in a
geometrically thick torus with a small opening angle. Noguchi et
al. (2009) found several buried AGN candidates with a very small
scattering fraction from the Second {\it XMM-Newton} Serendipitous
Source Catalogue ($2XMM$) using hardness ratios (HRs) and showed that
such type of AGNs tend to have a low relative
[\ion{O}{3}]$\lambda$5007 luminosity compared to the intrinsic X-ray
luminosity. This implies that the AGN with a small scattering fraction
could constitute the main class of optically elusive obscured AGNs.

In this paper, we construct a new sample of obscured AGNs covering a
broad range of scattering fractions from $2XMM$ Catalogue in the same
way as Noguchi et al. (2009), and investigate multi-wavelength
properties of a buried AGN in comparison with a classical type of AGN
having a large scattering fraction. In Section 2, we describe the
selection method of a new sample of obscured AGNs that covers a broad
range of scattering fractions. Our results of X-ray and optical  spectral analysis
are briefly presented in Section 3, and their multiwavelength
properties are discussed in Section 4. Finally, we summarize our
results in Section 5. We adopt ({\it H}$_{\rm 0}$, $\Omega_{\rm m}$,
$\Omega_{\rm \lambda}$) $=$ (70 km s$^{-1}$Mpc$^{-1}$, 0.3, 0.7)
throughout this paper.

\section{Sample Selection}
We used the $2XMM$ Catalogue produced by the {\it XMM-Newton} Survey
Science Centre, which contains $\sim$250,000 detections drawn from
$\sim$3500 {\it XMM-Newton} EPIC observations made between 2000 and
2007 (Watson et al. 2009).  The median flux in the full energy band
(0.2$-$12 keV) is $\sim$2.5$\times$10$^{-14}$ erg cm$^{-2}$ s$^{-1}$,
and about 20\% of the sources have total fluxes below
1$\times$10$^{-14}$ erg cm$^{-2}$ s$^{-1}$.

\begin{figure}[!b]
\centering
\includegraphics[angle=270,scale=.40]{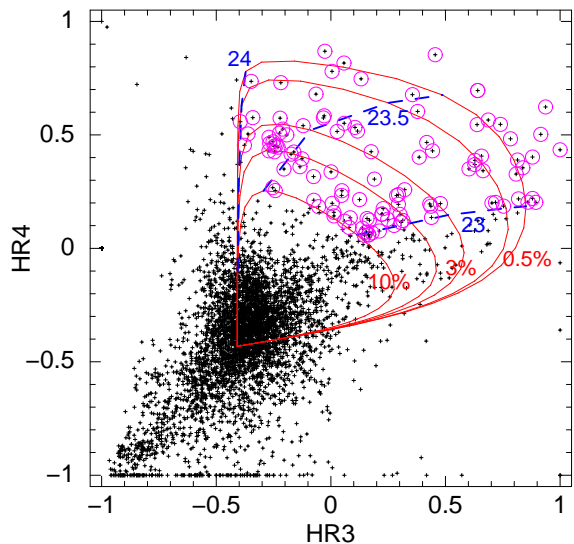}
\caption{Distribution of hardness ratio (HR) for the $2XMM$ Catalogue
  sources (crosses). Objects satisfying all of the following
  conditions are plotted; count rate in 0.2$-$12 keV $>$ 0.05 counts
  s$^{-1}$, $|b|$ $>$ 20$^\circ$, and HR error $\leq$ 0.2. Data points with
  circles are analyzed and 32 AGNs were selected as our sample from
  them.  Solid lines show the HRs expected for the scattering fraction
  of 10, 5, 3, 1, and 0.5\% from inside to outside. Dashed lines
  correspond to log$N$$_{\rm H}$ of 23, 23.5, and 24 cm$^{-2}$ from
  lower right to upper left.}
\label{figure:HR}
\end{figure}

We derived a sample that covers a wide range of scattering fractions
($\sim$0.1\%$-$10\%) from the $2XMM$ Catalogue using the HRs in the same
way as Noguchi et al. (2009) as briefly summarized below. We used HRs
defined as
\begin{displaymath}
 {\rm HR}3=\frac{{\rm CR}(2.0-4.5\ {\rm keV})-{\rm CR}(1.0-2.0\ {\rm
     keV})}{{\rm CR}(2.0-4.5\ {\rm keV})+{\rm CR}(1.0-2.0\ {\rm keV})}
\end{displaymath}
and 
\begin{displaymath}
 {\rm HR}4=\frac{{\rm CR}(4.5-12\ {\rm keV})-{\rm CR}(2.0-4.5\ {\rm
     keV})}{{\rm CR}(4.5-12\ {\rm keV})+{\rm CR}(2.0-4.5\ {\rm keV})},
\end{displaymath}
where CR(1.0$-$2.0 keV), CR(2.0$-$4.5 keV), and CR(4.5$-$12 keV) are
count rates in the 1.0$-$2.0, 2.0$-$4.5, and 4.5$-$12 keV bands,
respectively. Only 4627 sources, which have count rate for EPIC-pn in
0.2$-$12 keV $>$ 0.05 counts s$^{-1}$, high Galactic latitude ($|b|$ $>$
20$^\circ$), and error of HRs $\leq$ 0.2 at a 90\% confidence level,
were targeted in this selection. The values and errors of HR3 and HR4
given in the $2XMM$ Catalogue were calculated using count rates measured
by the {\tt emldetect} task in the {\it XMM-Newton} Science Analysis
System (SAS). In Figure \ref{figure:HR}, their HR3 and HR4 are
plotted. The five solid lines represent the scattering fractions of
10\%, 5\%, 3\%, 1\%, and 0.5\% from inside to outside. The three dashed lines
correspond to objects with log {\it N}$_{\rm H}$ (cm$^{-2}$) = 23,
23.5, and 24 from lower right to upper left. In calculating these
lines, we assumed an intrinsic spectrum of a power law with a photon
index of 1.9. The details on how to draw these lines are described in
Noguchi et al. (2009). Based on the solid line for 10\% and dashed
lines for log {\it N}$_{\rm H}$ = 23 and 24, we selected objects as
shown by circles in this figure as our sample for which log {\it
  N}$_{\rm H}$ and scattering fraction ($f_{\rm scat}$) are in the
range 23$-$24 and less than 10\%, respectively.

We excluded low-luminosity AGNs with the intrinsic 2$-$10 keV
luminosity lower than 10$^{41}$ erg s$^{-1}$ and Compton-thick AGNs
with column densities of log {\it N}$_{\rm H}$ $>$
1.5$\times$10$^{24}$ cm$^{-2}$ from our sample because of the
following reasons. Soft X-rays of low-luminosity AGNs would be
contributed by X-ray emission from hot gas and discrete sources in
their host galaxies, and direct emission of Compton-thick AGNs is
almost completely blocked below 10 keV. Therefore, scattering
fractions of these classes of objects cannot be calculated
correctly. We analyzed X-ray spectra obtained by {\it XMM-Newton} to
identify Compton-thick and low-luminosity AGNs.
If an X-ray continuum spectrum above 4 keV is fitted by a pure
reflection component
and the equivalent width of Fe-K$\alpha$ with respect to the
reflection component is greater than 1 keV (Matt et al. 1991), we
classified such objects as a Compton-thick AGN. The absorption
corrected 2$-$10 keV luminosities were calculated from the best-fit
models obtained in the same ways as discussed in Section 3.1, and
objects with the luminosity lower than 10$^{41}$ erg s$^{-1}$ were
regarded as a low-luminosity AGN.  Objects listed as a Seyfert 1, 1.5
or a star in NASA/IPAC Extragalactic Database (NED) were
excluded. Objects with a scattering fraction more than 10\%, which was
calculated from a formula described in Section 4.1, were also removed
from our sample. Furthermore, NGC 1052 was excluded from the sample
because this object shows complicated spectrum requiring a double
partial covering model, and a component absorbed by $N_{\rm H}$ $\sim$
2$\times$10$^{22}$ cm$^{-2}$ is relatively strong, which makes it
difficult to measure the scattering fraction.  We finally selected 32
objects as members of our sample, of which 10 objects are newly
selected AGNs and the others are included in Noguchi et al. (2009)
sample.

A summary of the newly selected AGNs and their properties is found in
Table \ref{table:sample}.  Most of the AGNs in our sample are at $z$ $<$
0.1 because we assumed $z$ = 0 in the simulation of AGN spectra used
to calculate the expected hardness ratios and our sample was selected
from bright sources (count rate in 0.2$-$12 keV $>$ 0.05 counts
s$^{-1}$). The Galactic column densities are calculated from 21 cm
measurements (Kalberla et al. 2005) using the {\tt nh} tool at the
NASA's High Energy Astrophysics Science Archive Research Center. Since
{\it XMM-Newton} results of some of the objects in our sample have
been published, references for them are also listed in Table
\ref{table:sample}.

{\tabletypesize{\scriptsize}
\begin{deluxetable*}{llcccccccc}
\setlength{\tabcolsep}{-3pt}
\tablewidth{0pt}
\tablecaption{Newly Selected AGNs}
\tablehead{
 \colhead{$2XMM$ Name}  &  
 \colhead{Other Name}  & 
 \colhead{Class}  &  
 \colhead{Reference$^a$}  &
 \colhead{Redshift}  & 
 \colhead{$N_{\rm H}$$^b$} & 
 \colhead{Start Date}  & 
 \colhead{Exposure$^c$}  & 
 \colhead{Count Rate$^d$} & 
 \colhead{Reference$^e$} \\
 \colhead{}   & 
 \colhead{}   &
 \colhead{}   &  
 \colhead{}   & 
 \colhead{}   & 
 \colhead{($10^{20}$ cm$^{-2}$)} &    
 \colhead{}    & 
 \colhead{(s)} & 
 \colhead{(counts s$^{-1}$)}&
 \colhead{} 
 }
 \startdata
  $2XMM$ J030030.5$-$112456   & MCG $-$02$-$08$-$039      &  Sy2    & 1       & 0.030     & 5.13 & 2006 Jan 23 & 5335  & 0.14  & \nodata\\
  $2XMM$ J031000.0+170559     & 3C 79                     &  Sy2   & 1       & 0.256     & 8.72 & 2004 Feb 14 & 9302  & 0.035 & 3\\
  $2XMM$ J033336.3$-$360825   & NGC 1365                  &  Sy1.8  & 1       & 0.005    & 1.34 & 2003 Jan 17 & 15142 & 0.66  & 4\\
  $2XMM$ J085537.7+781324     & NGC 2655                  &  Sy2    & 2       & 0.005     & 2.18 & 2005 Sep 04 & 3117  & 0.20  & 5\\
  $2XMM$ J104943.4+583750     & 2MASX J10494334+5837501   & \nodata & \nodata & 0.115     & 0.667& 2005 Oct 10 & 21908 & 0.044 & \nodata \\
  $2XMM$ J113240.2+525701     & Mrk 176                   &  Sy2    & 1       & 0.027     & 1.09 & 2004 May 02 & 9639  & 0.053 & 6\\
  $2XMM$ J134442.1+555312     & Mrk 273                   &  Sy2    & 1       & 0.038     & 0.890& 2002 May 07 & 18018 & 0.067 & 7\\
  $2XMM$ J135602.7+182218     & Mrk 463                   &  Sy2    & 1       & 0.051     & 2.03 & 2001 Dec 22 & 21081 & 0.12  &8, 9\\
  $2XMM$ J201657.3$-$704459   & IC 4970                   & \nodata & \nodata & 0.016     & 4.09 & 2002 Mar 31 & 19129 & 0.015 &\nodata\\
  $2XMM$ J224937.0$-$191627   & MCG $-$03$-$58$-$007      &  Sy2    & 1       & 0.0315     & 2.06 & 2005 May 09 & 2024  & 0.13  & \nodata
\enddata

\label{table:sample}
\tablenotetext{a}{References for optical classification.}
\tablenotetext{b}{Galactic column density by 21 cm measurement (Kalberla et al. 2005).}
\tablenotetext{c}{Cleaned exposure of EPIC-pn.}
\tablenotetext{d}{Count rate in the 0.4$-$10 keV band.}
\tablenotetext{e}{References for published {\it XMM-Newton} results.}

\tablerefs{(1)Veron-Cetty \& Veron 2006 (2) Ho et al. 1997; (3) Hardcastle et al. 2006; (4) Risaliti et al. 2005; (5) Akylas \& Georgantopoulos 2009; (6) Guainazzi et al. 2005; (7) Balestra et al. 2005; (8) Awaki et al. 2006; (9) Bianchi et al. 2008.}
\end{deluxetable*}
}

\section{Spectral Analysis}

\subsection{X-ray Spectra}
We analyzed X-ray spectra of newly selected sources to calculate the
scattering fraction. In the current analysis, the spectra were grouped
to a minimum of 15 counts per bin and the $\chi^2$ minimization
technique was used. The quoted errors are given at the 90\% confidence
level for one interesting parameter (i.e., $\Delta\chi^2$ = 2.71).  We
performed spectral fits for our sample in the 0.4$-$10 keV range with
various models by using XSPEC version 11.2.
First, all spectra were fitted with a baseline model consisting of
absorbed and unabsorbed power laws, along with a Gaussian
corresponding to an Fe-K$\alpha$ line. {\tt zpowerlw} and {\tt zgauss}
models in XSPEC were used to account for the power laws and the
Gaussian line, respectively.  All the components are modified by the
Galactic absorption. Our baseline model is shown as {\tt
  phabs*(zpowerlw + zphabs*(zpowerlw + zgauss))} in XSPEC, where {\tt
  zphabs} and {\tt phabs} are models corresponding to photoelectric
absorption by cold matter at the redshift of the source and in our
Galaxy, respectively. The photon indices ($\Gamma$) of the two power
laws were linked.

None of the spectra of the 10 objects were satisfactorily fitted with
the baseline model.  The spectrum of IC 4970 was explained by adding
another absorption for the less absorbed power law to the baseline
model.  For the other objects, we used a complex model, in which an
optically thin thermal plasma model ({\tt mekal} model in XSPEC; Mewe
et al. 1985, Kaastra. 1992, Liedahl et al. 1995) and/or a Compton
reflection model ({\tt pexrav} model in XSPEC) were added to the
baseline model.
For the {\tt mekal} model, the abundance was fixed at 0.5 solar, where
the solar abundance table by Anders and Grevesse (1989) was
assumed. For the {\tt pexrav} model, the inclination angle of the
reflector and the high-energy cutoff of the incident power law were
fixed at 60$^{\circ}$ (0$^\circ$ corresponds to face-on) and at 300
keV, respectively. The solar abundance table by Anders \& Grevesse
(1989) was assumed. {\tt rel\_refl} parameter was set to $-$1 to
produce a reflection component only. The normalization and the photon
index of {\tt pexrav} were assumed to be the same as that of the
heavily absorbed power law.  Since the reflection component of NGC
1365 required further absorption, we applied a {\tt zphabs} model to
the {\tt pexrav} model.  Furthermore, if needed, we added another {\tt
  mekal} component and/or Gaussian lines as many as required to model
the soft part of the spectrum.

Table \ref{table:model} shows the best-fit models. The spectra with
the models are shown in Figure \ref{figure:sp}.
The spectral
parameters for the models are summarized in Tables \ref{table:para1}$-$\ref{table:para3}. $\Gamma$ were distributed
between $\sim$1.5 and 2.0, which is the range of values for Seyfert 2s
(Smith \& Done 1996), although we fixed the value at 1.9 for some
objects since uncertainties of $\Gamma$ became large ($>$ 20\% at the
90\% level of confidence for one interesting parameter) or $\Gamma$
was far from a typical range (about 1.4$-$2.4), if $\Gamma$ was left
free.  The obtained $N_{\rm H}$ were in the range of 10$^{23-24}$
cm$^{-2}$. We calculated absorption-corrected luminosities for the
power-law components in 2$-$10 keV and of soft components in 0.5$-$2
keV using the best-fit model. The 0.5$-$2 keV luminosities were
calculated using all the components except for the heavily absorbed
power law corresponding to direct emission. The calculated
luminosities are summarized in Table \ref{table:data}. The
luminosities for our sample are in the range of Seyferts
($\sim$10$^{41-44}$ erg s$^{-1}$).

\begin{figure*}
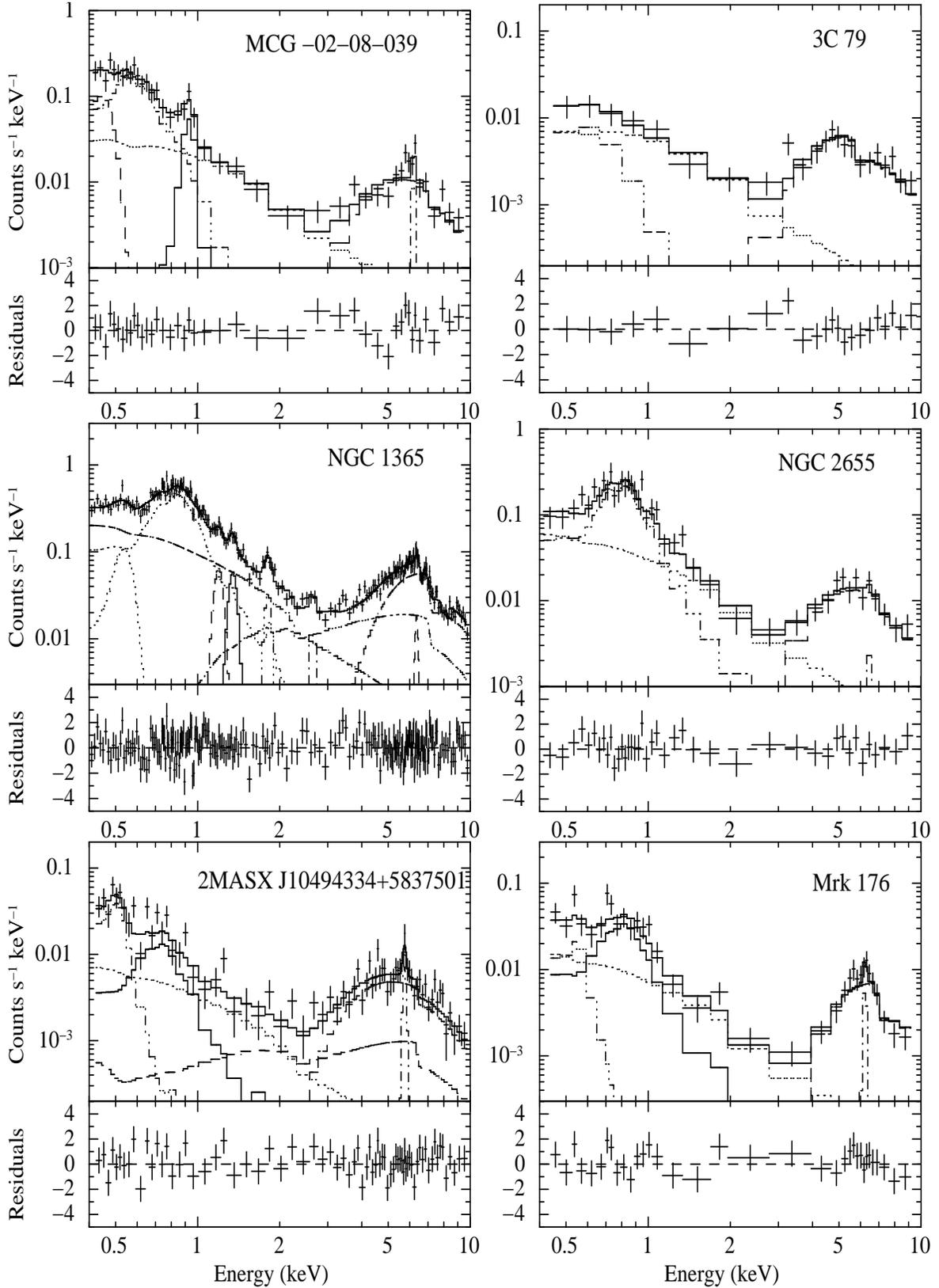

\centering
\includegraphics[width=7cm,height=8cm,clip,angle=270]{fig2a.eps}
\includegraphics[width=7cm,height=7.5cm,clip,angle=270]{fig2b.eps}
\includegraphics[width=7cm,height=8cm,clip,angle=270]{fig2c.eps}
\includegraphics[width=7cm,height=7.5cm,clip,angle=270]{fig2d.eps}
\includegraphics[width=7.5cm,height=8cm,clip,angle=270]{fig2e.eps}
\includegraphics[width=7.5cm,height=7.5cm,clip,angle=270]{fig2f.eps}
\caption{X-ray spectra ({\it upper panels}) and residuals in units of
  $\sigma$ ({\it lower panels}). Model components are shown with
  dashed, dotted, dot-dashed, and triple-dot-dashed lines.}
\label{figure:sp}
\end{figure*}

\begin{figure*}
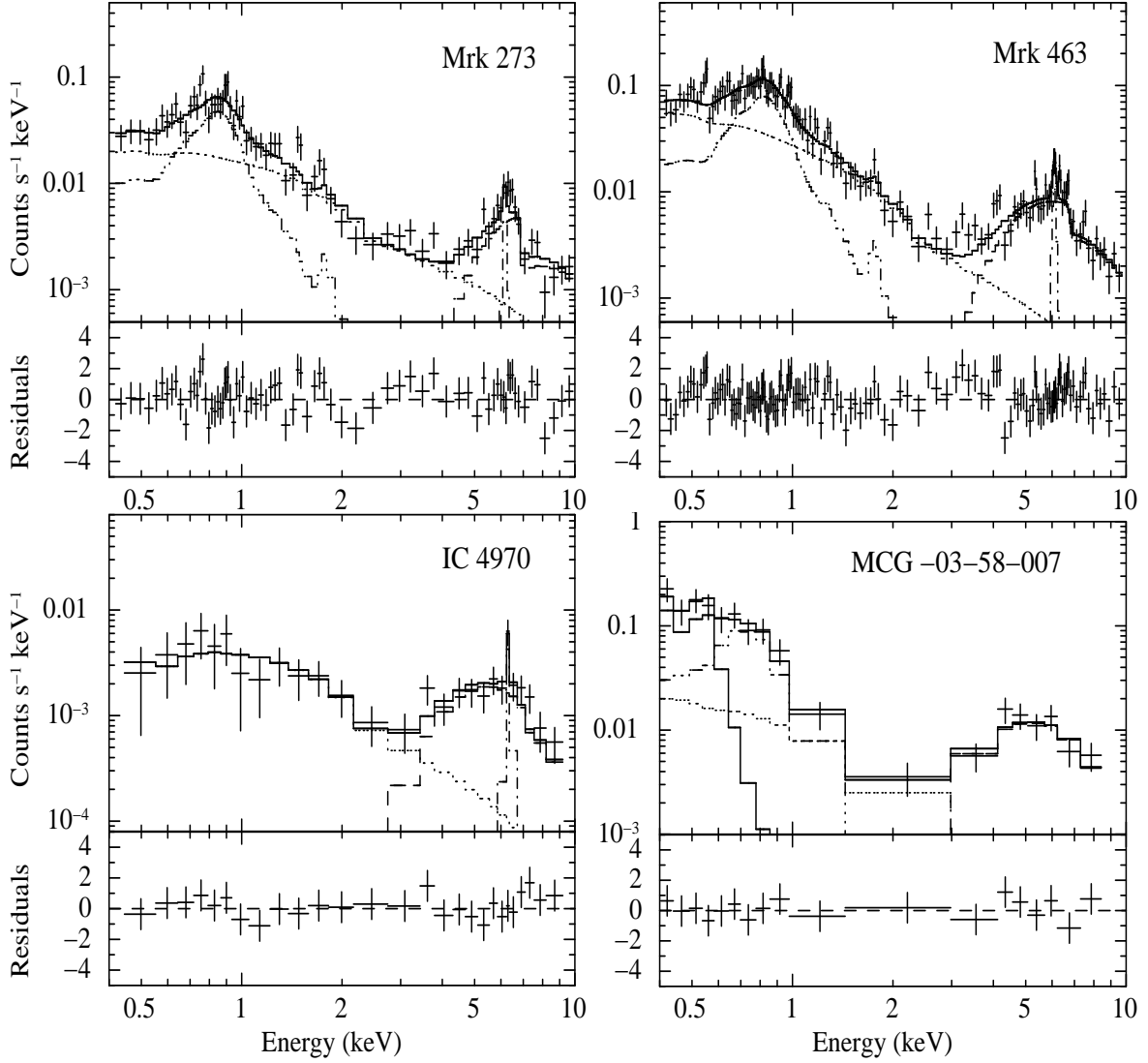

\figurenum{2}
\centering
\includegraphics[width=7cm,height=8cm,clip,angle=270]{fig2g.eps}
\includegraphics[width=7cm,height=7.5cm,clip,angle=270]{fig2h.eps}
\includegraphics[width=7.5cm,height=8cm,clip,angle=270]{fig2i.eps}
\includegraphics[width=7.5cm,height=7.5cm,clip,angle=270]{fig2j.eps}
\caption{Continued}
\end{figure*}

\begin{deluxetable}{ll}
\tabletypesize{\scriptsize}
\tablecaption{The Best-Fit Models of Newly Selected AGNs}
\tablewidth{0pt}
\tablehead{\colhead{Name}& \colhead{Model$^a$}
}
\startdata
  MCG $-$02$-$08$-$039  & BM + MEKAL + two lines  \\  
  3C 079                                 & BM + MEKAL \\ 
  NGC 1365                          & BM + MEKAL + nine lines + abs-Ref \\      
  NGC 2655                          & BM + MEKAL \\
  2MASX J10494334+5837501   & BM + two MEKAL + Ref \\
  Mrk 176                              & BM + MEKAL + abs-Ref \\  
  Mrk 273                               & BM + MEKAL \\
  Mrk 463                               & BM + MEKAL \\  
  IC 4970                               & BM$^b$\\ 
  MCG $-$03$-$58$-$007              & BM + two MEKAL 
\enddata
\tablenotetext{a}{BM: baseline model; PL: power law, MEKAL: thin thermal plasma model ({\tt mekal}), Ref: cold refection model ({\tt pexrav}), abs-Ref: absorbed ref. All components are absorbed by the Galactic absorption.}
\tablenotetext{b}{Additional absorption was applied to the less absorbed power law.}
\label{table:model}
\end{deluxetable}

{\tabletypesize{\scriptsize}
\begin{deluxetable*}{lccccccccc}
\tablecolumns{10}
\tablecaption{Spectral Parameters for Absorption, Power Law, and Gaussian in the Best-fit Models}
\setlength{\tabcolsep}{-3pt}
\tablewidth{0pt}
\tablehead{
\colhead{}\\    
 \colhead{Name}            & \colhead{$N_{\rm H}$}        & 
 \colhead{$\Gamma$}        & \colhead{$A_{\rm int}$$^a$}           & 
 \colhead{$E_{\rm line}$}   & \colhead{$\sigma$}          &  
 \colhead{EW}              &\colhead{$A_{\rm ga}$$^b$}    &
 \colhead{$A_{\rm scat}$$^c$}   &  \colhead{$\chi^2_{\rm \nu}$(dof)}        \\ 
 \colhead{}                &\colhead{($10^{22}$ cm$^{-2}$)}&
 \colhead{}                &                             &
 \colhead{(keV)}           & \colhead{(eV)}              & 
 \colhead{(eV)}            &                            &  
                           & \\
}               
 \startdata
  MCG $-$02$-$08$-$039            & 33$^{+13}_{-10}$     & 1.9(f)              & 0.92$^{+0.44}_{-0.28}$   & 6.435$^{+0.061}_{-0.066}$ & 10(f)     & 320$\pm200$  & 0.88$\pm0.54$              & 2.30$^{+0.47}_{-0.49}$ & 1.12(33)\\    
  3C 79                     & 48$^{+17}_{-13}$     & 1.9(f)              & 0.69$^{+0.30}_{-0.20}$   & 6.4(f)                  & 10(f)     & 0($<$110)      & 0($<$0.28)               & 1.18$\pm0.34$        & 0.93(17)\\   
  NGC 1365                  & 65.3$^{+2.2}_{-2.3}$ &2.056$\pm0.018$ & 12.4$^{+0.38}_{-0.28}$ & 6.390$^{+0.051}_{-0.055}$&0($<$124)&83$\pm40$&1.7$\pm1.1$ &10.09$^{+0.69}_{-0 .72}$&1.08(186)\\ 
  NGC 2655                  & 31$^{+10}_{-7}$      & 1.9(f)              & 1.11$^{+0.40}_{-0.30}$   & 6.4(f)                  & 10(f)       & 65($<$280)  & 0.21($<$0.90)              & 3.19$^{+0.85}_{-0.86}$ & 0.79(36)\\  
  2MASX J10494334+5837501   & 28.9$^{+7.8}_{-5.8}$ & 1.9(f)               & 0.367$^{+0.077}_{-0.062}$ & 6.403$^{+0.063}_{-0.072}$ & 60($<$195) & 300$^{+190}_{-150}$ & 0.36$^{+0.23}_{-0.18}$  & 0.40$^{+0.14}_{-0.13}$ & 1.14(65) \\
  Mrk 176                   & 62.4$^{+15.2}_{-7.5}$ & 1.9(f)              & 1.081$^{+0.17}_{-0.19}$  & 6.4(f)                  & 10(f)       & 200$\pm150$   & 0.66$\pm0.49$            & 0.83$^{+0.29}_{-0.23}$ & 1.11(27) \\
  Mrk 273                   & 82$^{+20}_{-16}$     & 1.59$^{+0.13}_{-0.16}$ & 0.52$^{+0.38}_{-0.23}$  & 6.4(f)                  & 10(f)      & 240$^{+190}_{-160}$ & 0.68$^{+0.53}_{-0.44}$  & 2.05$^{+0.36}_{-0.28}$ & 1.21(72) \\
  Mrk 463                   & 46.4$^{+6.5}_{-5.8}$  & 2.02$^{+0.13}_{-0.09}$ & 1.12$^{+0.46}_{-0.31}$  & 6.414$^{+0.040}_{-0.037}$ & 10(f)       & 250$\pm100$   & 0.70$\pm0.28$             & 3.21$^{+0.35}_{-0.36}$ & 1.02(136) \\
  IC 4970                   & 32$^{+15}_{-11}$     & 1.9(f)              & 0.34$^{+0.21}_{-0.13}$   & 6.410$^{+0.050}_{-0.062}$ &  0($<$100)     & 470$\pm300$   & 0.47$\pm0.30$       & 1.13$^{+0.24}_{-0.25}$ & 0.65(20)\\ 
  MCG $-$03$-$58$-$007            & 23.5$^{+9.6}_{-7.8}$ & 1.9(f)               & 0.78$^{+0.30}_{-0.28}$   &  6.4(f)                  & 10(f)      & 0($<$480)    & 0($<$1.14)                 & 1.12$^{+0.82}_{-0.97}$ & 0.68(10)
 \enddata
\tablecomments{Photon index of the power law with only Galactic absorption was assumed to be the same value as power law absorbed by cold matter at the redshift of the source. (f) indicates fixed parameter.}
\tablenotetext{a}{Normalization of the absorbed power law in units of $10^{-3}$ photons cm$^{-2}$ s$^{-1}$ at 1 keV.}
\tablenotetext{b}{Normalization of the Gaussian line in units of $10^{-5}$ photons cm$^{-2}$ s$^{-1}$ in the line.}
\tablenotetext{c}{Normalization of the less absorbed power law in units of $10^{-5}$ photons cm$^{-2}$ s$^{-1}$ at 1 keV.}
\label{table:para1}
\end{deluxetable*}
}

{\tabletypesize{\scriptsize}
\begin{deluxetable*}{lccccccc}
\setlength{\tabcolsep}{-3pt}
\tablecaption{Spectral Parameters for MEKAL and the Additional Absorptions in the Best-fit Models}
\tablewidth{0pt}
\tablehead{
\colhead{}\\    
               &&&&&&\multicolumn{1}{c}{Additional} \\
               &\multicolumn{2}{c}{MEKAL$^a$} &&\multicolumn{2}{c}{MEKAL$^a$}&\multicolumn{1}{c}{Absorption} \\ \cline{2-3}\cline{5-6}\\
\colhead{Name} &\colhead{\it kT}    & \colhead{$A_{\rm m}$$^b$}    &&\colhead{\it kT} & \colhead{$A_{\rm m}$$^b$} &\colhead{$N_{\rm H}$}          \\
               &\colhead{(keV)} &                       &&\colhead{(keV)} &                     &\colhead{($10^{22}$ cm$^{-2}$)}                    \\
}               
\startdata
  MCG $-$02$-$08$-$039          & 0.197$^{+0.021}_{-0.016}$ & 14.3$\pm3.3$      &&\nodata &\nodata  &\nodata \\
  3C 79                   & 0.35$^{+0.38}_{-0.16}$    & 1.14$^{+0.98}_{-0.76}$    &&\nodata &\nodata  &\nodata \\
  NGC 1365                 &0.635$\pm0.018$&22.26$^{+0.92}_{-0.94}$&&\nodata&\nodata&0.63$^{+0.54}_{-0.45}$$^c$\\ 
  NGC 2655                & 0.535$^{+0.096}_{-0.077}$ & 9.7$\pm1.4$          &&\nodata  &\nodata  &\nodata\\
  2MASX J10494334+5837501 & 0.54$^{+0.13}_{-0.14}$    & 0.612$^{+0.17}_{-0.16}$   && 0.081($<$0.085) & 28.8$^{+6.4}_{-6.2}$  &\nodata\\
  Mrk 176                 & 0.61$^{+0.12}_{-0.14}$    & 1.66$^{+0.45}_{-0.35}$    && 0.081($<$0.095) & 18$^{+10}_{-16}$  &\nodata\\
  Mrk 273                 & 0.695$^{+0.083}_{-0.057}$ & 3.00$^{+0.54}_{-0.59}$ &&\nodata &\nodata &\nodata \\
  Mrk 463                 & 0.664$\pm0.043$         & 4.15$^{+0.58}_{-0.56}$ &&\nodata &\nodata&\nodata \\
  IC 4970                 &\nodata&\nodata&&\nodata&\nodata&0.13$^{+0.11}_{-0.08}$$^d$\\ 
  MCG $-$03$-$58$-$007          & 0.36$^{+0.26}_{-0.10}$    & 5.2$^{+2.9}_{-2.1}$    && 0.081($<$0.11) & 125$^{+56}_{-100}$    &\nodata 
\enddata
\tablenotetext{a}{Metal abundances were fixed at 0.5 solar.}
\tablenotetext{b}{Normalization of {\tt mekal} in units of 10$^{-19}$/(4$\pi$({\it D}$_{\rm A}\times$(1+{\it z}))$^2$) $\int$ {\it n}$_{\rm e}$ {\it n}$_{\rm H}$ {\it dV}, where {\it D}$_{\rm A}$ is the angular size distance to the source (cm), {\it n}$_{\rm e}$ is the electron density (cm$^{-3}$), and {\it n}$_{\rm H}$ is the hydrogen density (cm$^{-3}$).}
\tablenotetext{c}{Absorption for a {\tt pexrav} model.}
\tablenotetext{d}{Absorption for a less absorbed power law.}
\label{table:para2}
\end{deluxetable*}
}

\newcounter{1}
\setcounter{1}{1}
\newcounter{4}
\setcounter{4}{4}
\newcounter{5}
\setcounter{5}{5}
\newcounter{6}
\setcounter{6}{6}
\newcounter{7}
\setcounter{7}{7}
\newcounter{8}
\setcounter{8}{8}
\newcounter{17}
\setcounter{17}{17}
\newcounter{9}
\setcounter{9}{9}
\newcounter{10}
\setcounter{10}{10}
\newcounter{11}
\setcounter{11}{11}
\newcounter{13}
\setcounter{13}{13}
\newcounter{16}
\setcounter{16}{16}
\newcounter{23}
\setcounter{23}{23}
\newcounter{25}
\setcounter{25}{25}
\newcounter{26}
\setcounter{26}{26}

{\tabletypesize{\scriptsize}
\begin{deluxetable}{lcccc}
\tablewidth{\columnwidth}
\tablecaption{Spectral Parameters for Additional Gaussians in the Best-fit Models}
\tablewidth{0pt}
\tablehead{
\colhead{}\\    
 \colhead{Name}       &\colhead{$E_{\rm line}$} &\colhead{$\sigma$}    &\colhead{$A_{\rm ga}$$^a$} &\colhead{Identification} \\  
                      &\colhead{(keV)}         & (eV)                &                           &                             \\
}               
\startdata
MCG $-$02$-$08$-$039                                & 0.467$^{+0.023}_{-0.043}$    &  10(f)              &3.5$^{+3.9}_{-1.7}$      & N \Roman{6} K$\alpha$, N \Roman{7} K$\alpha$  \\
                                                    & 0.959$\pm0.024$            &  10(f)              &0.70$\pm0.33$          & Ne \Roman{10} K$\alpha$ $?$, Ne \Roman{10} K$\alpha$ $?$   \\
                    \\
NGC 1365                                 &  0.544$\pm0.014$            &  10(f)             &2.42$\pm0.72$     & O \Roman{7} K$\alpha$\\
                                                    &  1.202$^{+0.019}_{-0.020}$    &  10(f)              &0.67$^{+0.23}_{-0.22}$    & Ne \Roman{5} Ly$\alpha$ \\
                                                    &  1.354$\pm0.19$            &  10(f)              &0.61$^{+0.14}_{-0.26}$           & Mg \Roman{4} K$\alpha$\\
                                                    &  1.820$^{+0.045}_{-0.028}$    &  10(f)              &0.37$\pm0.16$           &  Si \Roman{8}\\
                                                    &  2.65$^{+0.05}_{-0.12}$       &  10(f)             &0.29$\pm0.17$            & S \Roman{16} Ly$\alpha$\\
                                                    &  6.687$^{+0.037}_{-0.039}$    &  10(f)              &$-$2.66$^{+0.73}_{-0.74}$  & Fe \Roman{25} K$\alpha$\\
                                                    &  7.002$^{+0.040}_{-0.050}$    &  10(f)              &$-$1.48$\pm0.69$  & Fe \Roman{26} K$\alpha$\\
                                                    &  7.98$^{+0.09}_{-0.16}$       &  10(f)              &$-$1.6$\pm1.0$  & Fe \Roman{25} K$\beta$\\
                                                    &  8.287$^{+0.086}_{-0.069}$    &  10(f)              &$-$1.63$^{+0.94}_{-0.95}$  & Fe \Roman{26} K$\beta$                                      
\enddata
\tablecomments{(f) indicates fixed parameter.}
\tablenotetext{a}{Normalization of the Gaussian line in units of $10^{-5}$ photons cm$^{-2}$ s$^{-1}$.}
\label{table:para3}
\end{deluxetable}
}

{\tabletypesize{\scriptsize}
\begin{deluxetable*}{lccccccccccc}
\setlength{\tabcolsep}{-3pt}
\tablewidth{0pt}
\tablecaption{Multi-wavelength Properties of Our Sample}
\tablehead{
\colhead{Name} &
\colhead{Hard$^a$} & 
\colhead{Soft$^b$} & 
\colhead{$\log  L_{\rm FIR}$}  & 
\colhead{$f_{\rm 60}$/$f_{\rm 25}$$^c$}  &
\colhead{$\log L_{\rm [O III]}^{\rm int}$$^d$} & 
\colhead{Ref.$^e$} & 
\colhead{$M_{\rm BH}$$^f$} &
\colhead{Ref.$^g$} & 
\colhead{$L_{\rm bol}/L_{\rm Edd}$$^h$} &
\colhead{$f_{\rm scat}$$^i$}&
\colhead{$f_{\rm scat}^{\rm corr}$$^j$} 
}
\startdata
 \multicolumn{12}{l}{{\bf AGN in Noguchi et al. (2009)}}\\
  Mrk 348                   & 43.49 & 40.82 &  43.49   & 1.5     &  41.95    & 1       & 6.7     & 12      &  0.2    & 0.4 & 0.4     \\
  3C 33                     & 44.07 & 41.55 &  \nodata & \nodata &  42.52    & 2       & \nodata & \nodata & \nodata & 0.3 & \nodata \\
  2MASX J02281350$-$0315023 & 43.46 & 41.01 &  \nodata & \nodata &  \nodata  & \nodata & \nodata & \nodata & \nodata & 0.5 & \nodata \\
  NGC 1142                  & 43.51 & 41.25 &  44.77   & 8.4     &  41.87    & 3       & 8.2     & 12      & $-$1.3  & 0.8 & 0.3     \\
  3C 98                     & 43.01 & 41.00 &  \nodata & \nodata &  41.91    & 4       & \nodata & \nodata & \nodata & 2.0 & \nodata \\
  B2 0857+39                & 44.12 & 41.77 &  \nodata & \nodata &  \nodata  & \nodata & \nodata & \nodata & \nodata & 0.6 & \nodata \\
  IC 2461                   & 41.83 & 38.93 &  43.00   & \nodata &  40.85    & 5       & \nodata & \nodata & \nodata & 0.2 & \nodata \\
  2MASX J10335255+0044033   & 43.75 & 41.77 &  \nodata & \nodata &  43.62    & 6       & \nodata & \nodata & \nodata & 1.4 & \nodata \\
  MCG +08$-$21$-$065            & 42.65 & 39.95 &  43.51   & \nodata &  39.78    & 7       & \nodata & \nodata & \nodata & 0.3 & 0.1     \\
  NGC 4074                  & 42.88 & 40.88 &  \nodata & \nodata &  42.05    & 8       & 7.9     & 13      & $-$1.6  & 1.4 & \nodata \\ 
  NGC 4138                  & 41.21 & 38.88 &  \nodata & \nodata &  38.75    & 9       & 7.6     & 14      & $-$3.0  & 1.2 & \nodata \\
  NGC 4388                  & 42.89 & 40.63 &  43.94   & 2.9     &  41.77    & 1       & 7.0     & 13      & $-$0.7  & 1.4 & 0.8     \\
  NGC 4507                  & 43.09 & 41.19 &  43.81   & 3.1     &  41.69    & 1       & 7.5     & 12      & $-$1.0  & 3.1 & 2.9     \\
  ESO 506$-$G027              & 43.69 & 40.58 &  43.58   & 1.9     & \nodata$^k$ &\nodata   & \nodata & \nodata & \nodata & 0.2 & 0.1     \\
  2MASX J12544196$-$3019224 & 43.07 & 41.00 &  \nodata & \nodata &  41.60    & 5       & \nodata & \nodata & \nodata & 1.4 & \nodata \\
  NGC 4939                  & 42.33 & 40.31 &  43.59   & 5.4     &  41.43    & 1       & 7.6     & 15      & $-$1.8  & 1.3 & 0.8     \\
  ESO 383$-$G18               & 42.60 & 40.27 &  \nodata & 1.4     &  40.36    & 5       & \nodata & \nodata & \nodata & 1.3 & \nodata \\
  ESO 103$-$G035              & 43.33 & 40.35 &  43.54   & 1.0     &  41.65    & 1       & 7.0     & 15      & $-$0.3  & 0.1 & 0.1     \\
  IC 4995                   & 41.96 & 40.65 &  43.40   & 2.5     &  41.96    & 8       & 7.1     & 13      & $-$1.8  & 8.1 & 7.2     \\
  NGC 7070A                 & 41.75 & 39.33 &  42.40   & \nodata &  \nodata$^l$  & \nodata & 6.8     & 14      & $-$1.7  & 0.9 & 0.7     \\
  NGC 7172                  & 43.06 & 40.18 &  43.77   & 7.1     &  39.83    & 1       & 7.5     & 12      & $-$1.1  & 0.3 & 0.1     \\
  NGC 7319                  & 42.96 & 40.92 &  \nodata & \nodata &  41.44    & 1       & 7.2     & 13      & $-$0.9  & 1.3 & \nodata \\
 \hline
 \multicolumn{12}{l}{{\bf Newly selected AGN}} \\
  MCG $-$02$-$08$-$039            & 42.71 & 41.48 &  \nodata & 1.1     &  41.62    & 10      & 7.7     & 12      & $-$1.6  & 8.2 & \nodata \\  
  3C 079                    & 44.45 & 42.69 &  \nodata & 2.8     &  \nodata  & \nodata & \nodata & \nodata & \nodata & 2.4 & \nodata \\ 
  NGC 1365                  & 42.29 & 40.58 &  44.53   & 6.6     &  40.97    & 1       & \nodata & \nodata & \nodata & 2.1 & \nodata \\
  NGC 2655                  & 41.20 & 39.99 &  42.78   & 6.0     &  39.91    & 9       & 7.6     & 14      & $-$3.1  & 8.3 & 7.2     \\
  2MASX J10494334+5837501   & 43.49 & 41.83 &  \nodata & \nodata &  \nodata  & \nodata & \nodata & \nodata & \nodata & 3.0 & \nodata \\
  Mrk 176                   & 42.73 & 40.87 &  43.96   & 3.0     &  42.51    & 11      & 8.0     & 13      & $-$1.9  & 1.9 & 1.4     \\  
  Mrk 273                   & 42.90 & 41.42 &  45.53   & 9.5     &  42.45    & 1       & 8.1     & 12      & $-$1.8  & 7.3 & \nodata \\
  Mrk 463                   & 43.18 & 41.84 &  44.76   & 1.4     &  42.87    & 1       & \nodata & \nodata & \nodata & 5.3 & 4.4     \\  
  IC 4970                   & 41.71 & 39.96 &  \nodata & \nodata &  \nodata$^l$  & \nodata & \nodata & \nodata & \nodata & 2.4 & \nodata \\ 
  MCG $-$03$-$58$-$007            & 42.70 & 41.39 &  44.44   & 3.0     &  41.78    & 10      & \nodata & \nodata & \nodata & 6.6 & 5.1     
\enddata
\label{table:data}
\tablenotetext{a} {{\rm L}ogarithm of absorption-corrected 2$-$10 keV luminosity (erg s$^{-1}$). }
\tablenotetext{b} {{\rm L}ogarithm of absorption-corrected 0.5$-$2 keV luminosity of soft X-ray components (erg s$^{-1}$). }
\tablenotetext{c} {60 $\mu$m to 25 $\mu$m flux ratio. }
\tablenotetext{d} {{\rm L}ogarithm of [O III] luminosity (erg s$^{-1}$) corrected for reddening using Balmer decrement. }
\tablenotetext{e} {References for [O III] luminosity. }
\tablenotetext{f} {{\rm L}ogarithm of black hole mass calculated from stellar velocity dispersion. }
\tablenotetext{g} {References for stellar velocity dispersion. }
\tablenotetext{h} {{\rm L}ogarithm of Eddington ratio. }
\tablenotetext{i} {Scattering fraction (\%).}
\tablenotetext{j} {Scattering fraction corrected for the contribution of starburst (\%).}
\tablenotetext{k} {{\rm Lower} limit ($\log L_{\rm [O III]}^{\rm int}$$>43.23$ (erg s$^{-1}$)) is given in Landi et al. (2007).}
\tablenotetext{l} {[O III] lines are not detected.}

\tablerefs{(1) Bassani et al. 1999; (2) Yee \& Oke 1978; (3) Shu et
  al. 2007; (4) Costero \& Osterbrock 1977; (5) This work; (6) Dong et
  al. 2005; (7) Line flux measurement based on the Sloan Digital Sky
  Survey data at MPA/JHU (http://www.mpa-garching.mpg.de/SDSS/);
  Kauffmann et al. 2003; (8) Polletta et al. 1996; (9) Ho et al. 1997;
 (10) de Grijp et al. 1992; (11) Mulchaey et
  al. 1994; (12) Garcia-Rissmann et al. 2005; (13) Nelson \& Whittle
  1995; (14) McElroy 1995; (15) Cid Fernandes et al. 2004}
\end{deluxetable*}
}

\subsection{Optical Spectra}
\begin{figure*}
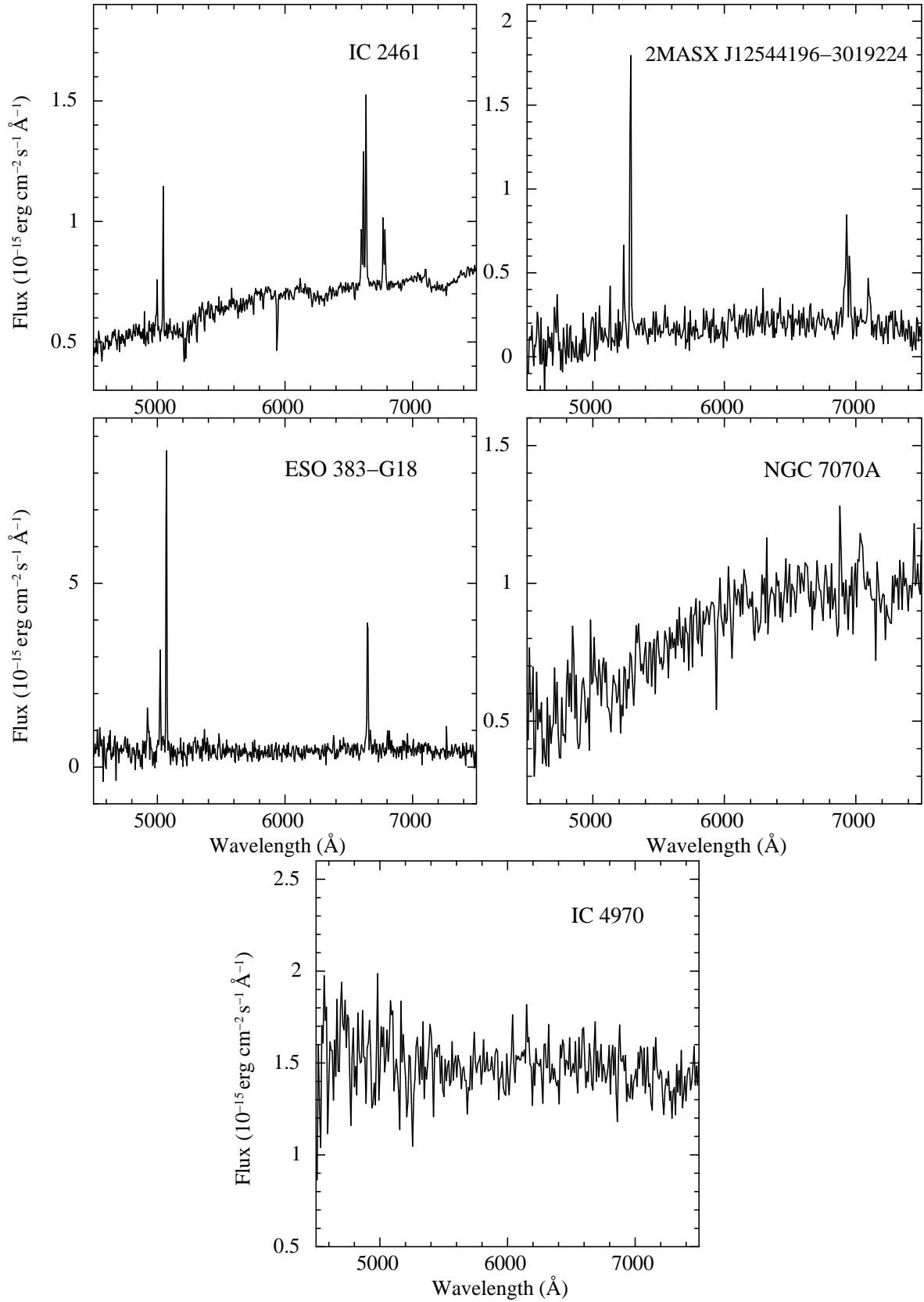

\centering
\includegraphics[width=7cm,height=8cm,clip,angle=270]{fig3a.eps}
\includegraphics[width=7cm,height=7.5cm,clip,angle=270]{fig3b.eps}
\includegraphics[width=7.5cm,height=8cm,clip,angle=270]{fig3c.eps}
\includegraphics[width=7.5cm,height=7.5cm,clip,angle=270]{fig3d.eps}
\includegraphics[width=7.5cm,height=8cm,clip,angle=270]{fig3e.eps}
\caption{Optical spectra of IC 2461, 2MASX J12544196$-$3019224, 
ESO 383$-$G18, NGC 7070A, and IC 4970 in the 4500-7500 \AA \ wavelength range.}
\label{figure:sp_opt}
\end{figure*}

Optical spectroscopic observations of four targets in our sample,
2MASX~J12544196--3019224, IC~4970, NGC~7070A, and ESO~383--G18, were
performed during several nights between 2008 August 9 and 11 by using
the South African Astronomical Observatory (SAAO) 1.9 m telescope with the
Cassegrain spectrograph.  The grating six, which has a spectral range
of about 3500$-$5300 \AA ~at a resolution around 4 \AA, was used with
$\sim$ 2 arcsec slit placed on the center of each galaxy, for a total
integration time ranging from 750 to 3600 s. The data reduction and
analysis was made in a standard manner with the IRAF package to derive
the flux calibrated spectra. To obtain the sensitivity curve, we
fitted the observed spectral energy distribution (SED) of standard stars
with low-order polynomials. We detected [\ion{O}{3}]$\lambda$5007 as
well as H$\alpha$ and H$\beta$ narrow lines from
2MASX~J12544196--3019224 and ESO~383--G18. Their narrow lines and line
intensity ratios indicate that 2MASX~J12544196--3019224 and
ESO~383--G18 are a Seyfert 2 and \ion{H}{2} nucleus, respectively.
For IC~4970 and NGC~7070A, we derived the 90\% upper limits on the
[\ion{O}{3}] intensity, based on an estimated rms noise in the
continuum flux around the corresponding wavelength. 
The fluxes were corrected for the slit loss, estimated from
the image extent of the target or a nearby point-like source projected
onto the spatial direction. We also analyzed
calibrated optical spectrum of IC 2461 taken from the Sloan Digital
Sky Survey (SDSS) Data Release 7. The
narrow lines and line intensity ratios indicate a Seyfert 2.  The line
fluxes and the spectra of the five objects are shown in Table
\ref{table:optical} and Figure \ref{figure:sp_opt}, respectively.

\begin{deluxetable}{lcccc}
\tabletypesize{\scriptsize}
\tablecaption{[O III]$\lambda$5007, H$\alpha$, and H$\beta$ Fluxes}
\tablewidth{0pt}
\tablehead{\colhead{Name}& \colhead{[O III]} & \colhead{H$\alpha$} & \colhead{H$\beta$} & \colhead{Class} 
}
\startdata
IC 2461                     &  4.2    & 3.9       & 0.50   & Sy2 \\
2MASX J12544196$-$3019224   &  4.1    & 1.9      & 0.58   & Sy2 \\
ESO 383$-$G18               &  14    & 6.5       & 2.8    & H II \\
NGC 7070A                   & $<$1  & \nodata   &\nodata & \nodata \\
IC 4970                     & $<$2  & \nodata   &\nodata & \nodata
\enddata
\tablecomments{Fluxes are in units of 10$^{-14}$ erg cm$^{-2}$ s$^{-1}$. Upper limits are at a 90\% confidence level.}
\label{table:optical}
\end{deluxetable}

\newpage
\section{Discussion}
\subsection{Scattering Fraction}
We first calculated a scattering fraction ({\it f}$_{\rm scat}$) for our sample 
consisting 32 AGNs with the equation
\begin{displaymath}
 {\it f}_{\rm scat}=\frac{L_{0.5-2}^{\rm soft}}{L_{0.5-2}^{\rm int}} , 
\end{displaymath}
where $L_{0.5-2}^{\rm int}$ and $L_{\rm 0.5-2}^{\rm soft}$ are
absorption corrected fluxes in the 0.5$-$2 keV band for the absorbed
power law and all the components except for the heavily absorbed power
law, respectively. The calculated values are shown in Table
\ref{table:data}.  Various components other than scattered emission,
however, may contribute to the soft X-rays in obscured AGNs such as
thermal emission originated from hot plasma collisionally heated by
starburst activity. In fact, high-resolution images available with
{\it Chandra} show that the soft X-ray emission is dominated by the
starburst component in some cases (NGC 4945, Schurch et al. 2002),
although the {\it XMM-Newton} Reflection Grating Spectrometer (RGS)
high-resolution spectra show evidence that the soft X-ray emission
appears to be dominated by AGN emission for most Seyfert 2 galaxies
with high-quality data (Guainazzi \& Bianchi 2007). Therefore, the
value calculated from the above equation is regarded as an upper limit
on the scattering fraction.

\begin{figure}[!htb]
\centering
\includegraphics[angle=270,scale=.40]{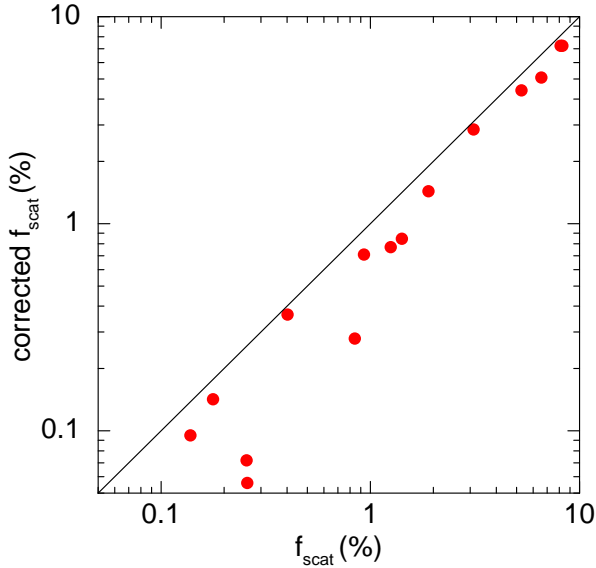}
\caption{Comparison of scattering fractions ($f_{\rm scat}$) with corrected scattering
  fractions ($f_{\rm scat}^{\rm corr}$) calculated by subtracting the starburst contribution
  estimated from FIR luminosity. The solid line represents $f_{\rm
    scat}$$=$$f_{\rm scat}^{\rm corr}$. }
\label{figure:fs}
\end{figure}

We examined the starburst contribution to the soft X-rays for our
sample using far-infrared luminosities ({\it L}$_{\rm FIR}$), and
calculated scattering fractions by subtracting the estimated
contribution. {\it L}$_{\rm FIR}$ is often used to estimate starburst
activity although we cannot rule out the contribution of AGNs and host
galaxies to {\it L}$_{\rm FIR}$. We calculated {\it L}$_{\rm FIR}$ for
our sample using the formula defined in Helou et al. (1985), based on
flux densities at 60 $\mu$m and 100 $\mu$m. We collected infrared
fluxes (60 $\mu$m and 100 $\mu$m) measured with {\it Infrared
  Astronomical Satellite} ({\it IRAS}) for 18 of 32 objects from NED.
{\it IRAS} fluxes for the rest of the 14 objects are not available.
Observations with upper limits were not used. {\it L}$_{\rm FIR}$ for
our sample is shown in Table \ref{table:data}.  Ranalli et al. (2003)
show that there is a strong correlation between {\it L}$_{\rm FIR}$
and $L_{\rm 0.5-2}^{\rm soft}$ for starburst and normal galaxies as
expressed by
 \begin{displaymath}
 {\rm log}{\it L}_{\rm 0.5-2}^{\rm soft} = {\rm log}{\it L}_{\rm
   FIR}^{\rm int} - 3.70.
 \end{displaymath}
We estimated a 0.5$-$2 keV luminosity created by starburst ({\it
  L}$_{0.5-2}^{\rm SB}$) using this correlation and calculated the
scattering fraction as
\begin{displaymath}
 {\it f}_{\rm scat}^{\rm corr}=\frac{L_{0.5-2}^{\rm soft} - {\it
     L}_{0.5-2}^{\rm SB}}{L_{0.5-2}^{\rm int}}.
\end{displaymath}
The calculated values are shown in Table \ref{table:data}. In Figure
\ref{figure:fs}, $f_{\rm scat}^{\rm corr}$ values are compared with
$f_{\rm scat}$.  IC 2461, NGC 1365, and Mrk 273 were not plotted in
this figure since the values of $L_{0.5-2}^{\rm soft}$ $-$
$L_{0.5-2}^{\rm SB}$ are negative. This means that starburst
contribution to their soft X-rays is substantial.  In fact, recent
{\it XMM-Newton} RGS spectra of NGC 1365 show soft X-ray emission
dominated by collisionally ionized plasma (Guainazzi et al. 2009).
Figure \ref{figure:fs} shows a relatively tight correlation between
$f_{\rm scat}^{ \rm corr}$ and $f_{\rm scat}$, although the values of
$f_{\rm scat}^{\rm corr}$ are lower than $f_{\rm scat}$ and the
scatter for objects with a small $f_{\rm scat}$ is larger than for a
large $f_{\rm scat}$. The tight correlation, particularly at large
$f_{\rm scat}$, is consistent with the idea that the soft X-rays
dominated by emission from photoionized gas are likely to be a common
characteristic of Seyfert 2 galaxies. Therefore, in the following
discussions, we use $f_{\rm scat}$ since the values can be calculated
for all objects in our sample.

From our spectral analysis combined with our previous results, we
found that $f_{\rm scat}$ for our sample are in the range of
$\sim$0.1\%$-$10\%. In particular, those of eight objects are very small
($<$ 0.5\%) as reported in Noguchi et al. (2009), whereas a typical
value for Seyfert 2s previously studied is about 3\% (Bianchi \&
Guainazzi 2007; Turner et al. 1997). The scattering fraction is
proportional to both the solid angle subtended by the scattering
electrons and a scattering optical depth.  Thus, such a very small
$f_{\rm scat}$ implies that they would be a new type of AGN buried in
a very geometrically thick torus with a small opening angle (Ueda et
al. 2007; Winter et al. 2008, 2009; Eguchi et al. 2009; Noguchi et
al. 2009) on the assumption that the optical depth of scatterer is
nearly constant among the objects.  Our sample that covers a broad
range of ${\it f}_{\rm scat}$ allows us to investigate properties of
buried AGNs with a small ${\it f}_{\rm scat}$ by comparing with those
of a classical type of Seyfert 2s with a large ${\it f}_{\rm scat}$.

\subsection{Geometry of Obscuring Material}
\begin{figure}[!htb]
\centering
\includegraphics[angle=270,scale=.40]{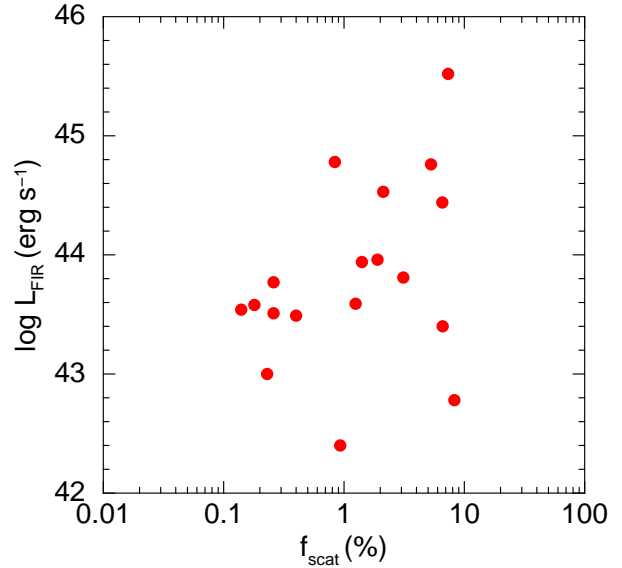}
\caption{Far infrared luminosity ($L_{\rm FIR}$) plotted against scattering fraction. 
}
\label{figure:FIR}
\end{figure}

Fabian et al. (1998) proposed a model in which low-luminosity AGNs
were obscured by starburst in the inner 100 pc of a central massive
black hole and suggested that supernovae from nuclear starburst input
the energy to the circumnuclear gas to create a torus-like structure.
This suggestion was investigated in more detail using
three-dimensional hydrodynamic simulations (Wada \& Norman 2002,
2007). Seyfert galaxies with nuclear starburst within a few hundred pc
from the center were indeed found observationally (Imanishi
2002, 2003; Rodriguez-Ardila \& Viegas 2003).  Therefore, nuclear
starburst is considered as a key factor to keep the shape of a
torus-like absorber either theoretically or observationally. If this
is the case, the scale height of the torus would expand with an
increase of the star-forming activity, and starburst activity in
objects with a small $f_{\rm scat}$ should be stronger than those in
objects with a large $f_{\rm scat}$. To confirm this expectation, we
investigated a relation between $f_{\rm scat}$ and {\it L}$_{\rm FIR}$
as shown in Figure \ref{figure:FIR}. We found no significant
correlation between them, in contrast to our expectation that there is an
anti-correlation.
This means that the scale height of the torus seems
not to be primarily determined by starburst activity and there are
other contributing factors to support the shape of a torus-like
absorber.

The absence of a correlation between $f_{\rm scat}$ and $L_{\rm FIR}$
might in part be due to a selection bias against infrared luminous
objects such as ultraluminous infrared galaxies (ULIRGs), which have
most of their luminosity emerging in the infrared with $L$(8$-$1000
$\mu$m) $>$ 3$\times$10$^{45}$ erg s$^{-1}$ and contain a
substantially large amount of gas and dust in their nuclei compared
with classical Seyferts (Sanders \& Mirabel 1996). Imanishi et
al. (2007) found buried AGNs signatures in a significant fraction of
ULIRGs using {\it Spitzer} Infrared Spectrograph even though they are
optically classified as non-Seyferts, and suggested that AGNs in ULIRGs
are almost fully buried in surrounding gas and dust, for which $f_{\rm
  scat}$$\sim$ 0 is expected. In our sample, however, only Mrk 273 and
Mrk 463 are classified as a ULIRG. The paucity of infrared luminous
objects in our sample is possibly due to our selection criteria since
many of ULIRGs are heavily obscured by Compton-thick matter $>$
10$^{24}$ cm$^{-2}$ (Risaliti et al. 2000; Teng et al. 2009). If we
had plotted them in Figure \ref{figure:FIR}, they would be located in
the upper left portion since they have {\it f}$_{\rm scat}$$\sim$ 0
and are bright in the infrared, and the anti-correlation may appear in
Figure \ref{figure:FIR} as expected.



\subsection{Infrared Color}
\begin{figure}[!htb]
\centering
\includegraphics[angle=270,scale=.40]{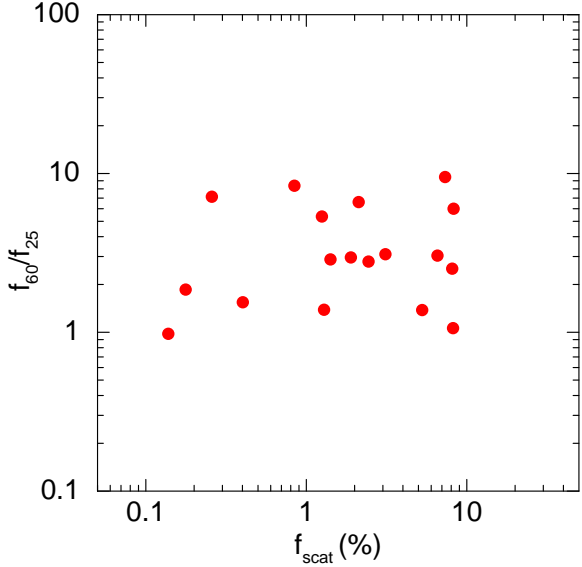}
\caption{Infrared color $f_{\rm 60}$/$f_{\rm 25}$ plotted against
  scattering fraction. 
}
\label{figure:color}
\end{figure}

Warm infrared colors of AGNs are usually explained as emission from
hot dust in the vicinity of a nucleus. For a buried AGN, the cooler
component from the outer region of the torus dominates, since a
geometrically thick torus almost covers the center region unless the 
viewing angle is near face-on. Hence, we
predict that buried AGNs with a small scattering fraction show cooler
infrared colors. 60 $\mu$m to 25 $\mu$m flux ratios ($f_{\rm 60}$/$f_{\rm
  25}$) for our sample are shown in Table \ref{table:data}.  Figure
\ref{figure:color} compares $f_{\rm scat}$ with $f_{\rm 60}$/$f_{\rm
  25}$ and shows that there is no correlation between them. This
result is inconsistent with above prediction. However, this may be
explained by the effect of the difference of inclination of the torus to
our line of sight because $f_{\rm 60}$/$f_{\rm 25}$ depends also on
the viewing angle (Heisler et al. 1997).  If we view a Seyfert galaxy
nearly face-on, the value of $f_{\rm 60}$/$f_{\rm 25}$ becomes smaller
because infrared emission from the hot dust near the nucleus can be
seen.  On the other hand, an edge-on view makes $f_{\rm 60}$/$f_{\rm
  25}$ larger since the hot dust is obscured and cooler component from
the outer part dominates.  Expected infrared SEDs are calculated by
several authors for smooth (Pier \& Krolik 1992, 1993; Granato
\& Danese 1994) and clumpy (Nenkova el al. 2008) distribution of
torus. According to these calculations, objects with small $f_{\rm
  scat}$ may be viewed from a face-on angle, in spite of the fact that
their absorption column densities along the line of sight is large
(10$^{23-24}$ cm$^{-2}$). A similar situation is found in {\it Swift}
BAT-selected AGNs (Ueda et al. 2007).  If we observe them from an
edge-on side, their direct emission would be completely blocked by
Compton-thick matter ($\gg$ 10$^{24}$ cm$^{-2}$). Thus, if our results
are explained in terms of face-on geometry, this implies the existence of a
yet to be discovered very Compton-thick AGN.
 
\subsection{Scattering Fraction and [\ion{O}{3}] }
If a scattering fraction reflects the opening angle of the torus,
objects with a small $f_{\rm scat}$ should have also a small narrow-line
 region (NLR), which is considered to exist in the opening region
of the torus. We predicted that their [\ion{O}{3}] $\lambda$5007
emission luminosities produced in the NLR are weaker than those for
objects with a large $f_{\rm scat}$ at a given X-ray luminosity (Ueda
et al. 2007; Noguchi et al. 2009). To test this prediction, Noguchi et
al. (2009) compared ratios of reddening corrected [\ion{O}{3}] to
intrinsic 2$-$10 keV luminosities ({\it L}$_{\rm [O\ III]}^{\rm
  int}$/{\it L}$_{\rm 2-10}^{\rm int}$) for a sample of the buried AGN
consisting of 22 objects with a small $f_{\rm scat}$ with those for a
large sample of Seyfert2s compiled by Bassani et al. (1999), and found
that there is a clear difference in their distributions, with the buried
AGNs having smaller ratios of {\it L}$_{\rm [O\ III]}^{\rm int}$/{\it
  L}$_{\rm 2-10}^{\rm int}$. This result is in good agreement with
the above prediction. In this subsection, to confirm the result we
examine whether there is a similar relationship between {\it L}$_{\rm
  [O\ III]}^{\rm int}$/{\it L}$_{\rm 2-10}^{\rm int}$ and $f_{\rm
  scat}$ using our sample, which covers a broad range of $f_{\rm
  scat}$.

We used [\ion{O}{3}] luminosities collected from the literature as
shown in Table \ref{table:data}. These values are corrected for the
extinction by using the relation
\begin{displaymath}
 {\it L}_{\rm [O\ III]}^{\rm int}={\it L}_{\rm [O\ III]}^{\rm obs}\biggl[\frac{{\rm
       H}\alpha/{\rm H}\beta}{({\rm H}\alpha/{\rm
       H}\beta)_0}\biggr]^{2.94},
\end{displaymath}
assuming an intrinsic Balmer decrement (H$\alpha$/H$\beta$)$_0$ = 3.0,
where $L_{[{\rm O III}]}^{\rm obs}$ and {\rm H}$\alpha/${\rm H}$\beta$
are an observed [\ion{O}{3}] luminosity and a ratio between observed
{\rm H}$\alpha$ and {\rm H}$\beta$ line fluxes, respectively (Bassani
et al. 1999). The calculated values are shown in Table \ref{table:data}.

\begin{figure}[!t]
\centering
\includegraphics[angle=270,scale=.40]{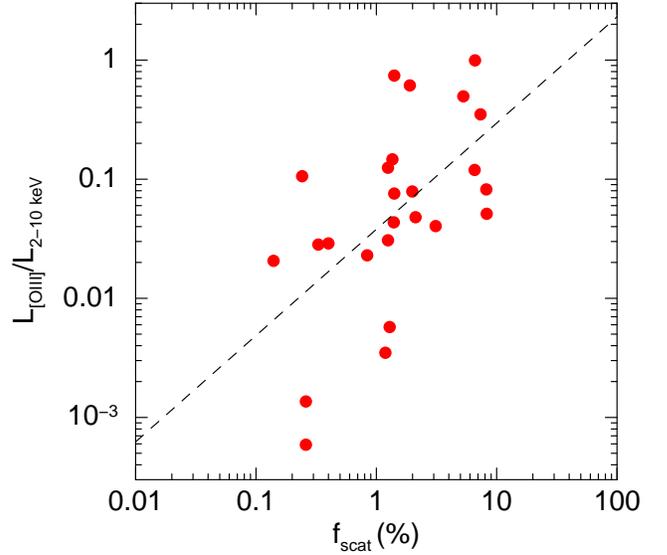}
\caption{Ratio of intrinsic 2$-$10 keV to
  reddening corrected [\ion{O}{3}] line luminosities plotted against
  scattering fraction. The dashed line is a linear fit to the data.
  Spearman's and Kendall's rank correlation coefficients are $\rho$ $=$ 0.62 ({\it p} $=$ 0.0011) and
$\tau$ $=$ 0.45 ({\it p} $=$ 0.0014), respectively. 
}
\label{figure:opt}
\end{figure}

In Figure \ref{figure:opt}, we plot {\it L}$_{\rm [O\ III]}^{\rm
  int}$/{\it L}$_{\rm 2-10}^{\rm int}$ against $f_{\rm scat}$. As
expected, we find there is a positive correlation between them. The
best-fit liner line (dashed line in Figure \ref{figure:opt}) is given
as
\begin{displaymath}
 {\rm log} \frac{{\it L}_{\rm [O\ III]}^{\rm int}}{{\it L}_{\rm
     2-10}^{\rm int}} = (0.89\pm0.43) {\rm log}\ {\it f}_{\rm scat} -
 (1.42\pm0.23).
\end{displaymath}
We computed a Spearman's rank correlation coefficient ($\rho$) and
Kendall's rank correlation coefficient ($\tau$) to determine a level
of significance, and found $\rho$ $=$ 0.62 ({\it p} $=$ 0.0011) and
$\tau$ $=$ 0.45 ({\it p} $=$ 0.0014), where {\it p} is null hypothesis
probability. This correlation clearly supports the result obtained by
Noguchi et al. (2009) showing {\it L}$_{\rm [O\ III]}^{\rm int}$/{\it
  L}$_{\rm 2-10}^{\rm int}$ for buried AGNs tend to be smaller than
other Seyfert 2s. Therefore, as suggested in Noguchi et al. (2009),
estimation of intrinsic luminosities of buried AGNs based on
[\ion{O}{3}] luminosities would result in large uncertainties, and
surveys using optical emission lines could be subject to biases
against buried AGNs.

\subsection{Black Hole Mass and Eddington Ratio}
\begin{figure}[!htb]
\centering
\includegraphics[angle=270,scale=.40]{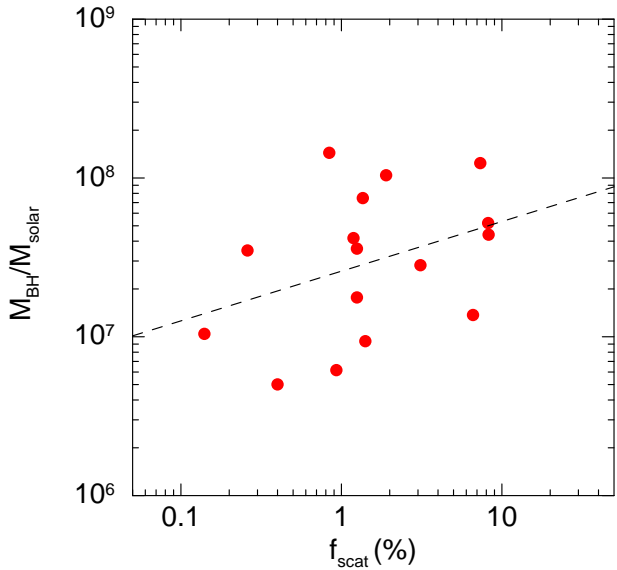}
\caption{Black hole mass plotted against scattering fraction. The
  dashed line is a linear fit to the data. Spearman's and Kendall's rank correlation 
  coefficients are $\rho$ $=$ 0.36 ({\it p} $=$ 0.17) and $\tau$ $=$ 0.25 ({\it p} $=$ 0.19), 
  respectively. 
}
\label{figure:MBH}
\end{figure}

The black hole mass ($M_{\rm BH}$) is one of the most important pieces
of information to represent properties of an AGN, and we compare
$f_{\rm scat}$ with $M_{\rm BH}$. Black hole masses for our sample
were estimated from the relation found by Tremaine et al. (2002)
\begin{displaymath}
M_{\rm BH}=10^{8.13}\times\biggl[\frac{\sigma_{\ast}}{200\ {\rm
      km\ s^{-1}}}\biggr]^{4.02}M_{\sun},
\end{displaymath}
where $\sigma_{\ast}$ is a stellar velocity dispersion. The values of
$\sigma_{\ast}$ for 16 among 32 objects are collected from the
literature as shown in Table \ref{table:data}.  The black hole masses
for our sample are in the range 6.7 $\leq$ log($M_{\rm BH}/M_{\sun}$)
$\leq$ 8.2 (Table \ref{table:data}).  In Figure \ref{figure:MBH}, we
plot $M_{\rm BH}$ against $f_{\rm scat}$ and found only a hint of very
weak positive correlation between them ($\rho$ $=$ 0.36 ($p$ $=$ 0.17)
and $\tau$ $=$ 0.25 ($p$ $=$ 0.19)). If this trend is real, it is
consistent with an X-ray absorption model for a compton-thin AGN
described in Lamastra et al. (2006), in which a higher black hole mass
leads to a larger opening angle.

\begin{figure}[!htb]
\centering
\includegraphics[angle=270,scale=.40]{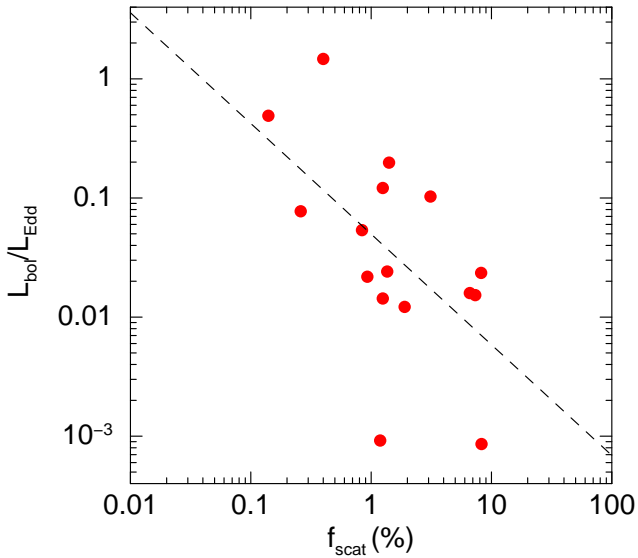}
\caption{Eddington ratio plotted against scattering fraction. The
  dashed line is a linear fit to the data. Spearman's and Kendall's rank correlation 
  coefficients are $\rho$ $=$ $-$0.54 ({\it p} $=$ 0.032) and $\tau$ $=$ $-$0.40 ({\it p} $=$ 0.033), 
  respectively.
}
\label{figure:Edd}
\end{figure}

The Eddington ratio ($L_{\rm bol}/L_{\rm Edd}$) is also an important
parameter in studies of accretion and evolution of AGNs. Eddington
luminosities ($L_{\rm Edd}$) are calculated from the black hole mass
as $L_{\rm Edd}$=1.26$\times$10$^{38}$($M_{\rm
  BH}/M_{\sun}$). Bolometric luminosities ($L_{\rm bol}$) are
calculated as $L_{\rm bol}=$30$\times$$L_{\rm 2-10}^{\rm int}$, where
a typical bolometric correction factor for luminous AGNs ($\sim$ 30) is
applied (Elvis et al. 1994; Risaliti \& Elvis 2004; Vasudevan \&
Fabian 2007) and $L_{\rm 2-10}$ is an intrinsic luminosity in the 2$-$
10 keV band. We note that [\ion{O}{3}] luminosities may not be a good
estimator of intrinsic luminosities as discussed in Section 4.2,
although they are often used to derive bolometric luminosities
(Heckman et al. 2004). The values of $L_{\rm bol}/L_{\rm Edd}$ are in
the range 10$^{-4}$ $<$ $L_{\rm bol}/L_{\rm Edd}$ $<$ 2. In Figure
\ref{figure:Edd}, the $L_{\rm bol}/L_{\rm Edd}$ ratios are plotted
against $f_{\rm scat}$, and we find an anti-correlation between them
($\rho$ $=$ $-$0.54 ($p$ $=$ 0.032) and $\tau$ $=$ $-$0.40 ($p$ $=$
0.033)). This correlation could indicate that buried AGNs are rapidly
growing compared with AGNs surrounded by the torus with a large opening
part. {\it f}$_{\rm scat}$ could be an indicator of the growing phase
of black holes and a useful parameter to understand the evolution of
the active nucleus.


\subsection{Host Galaxy}
\begin{figure*}[!htb]
\centering
\includegraphics[scale=0.85]{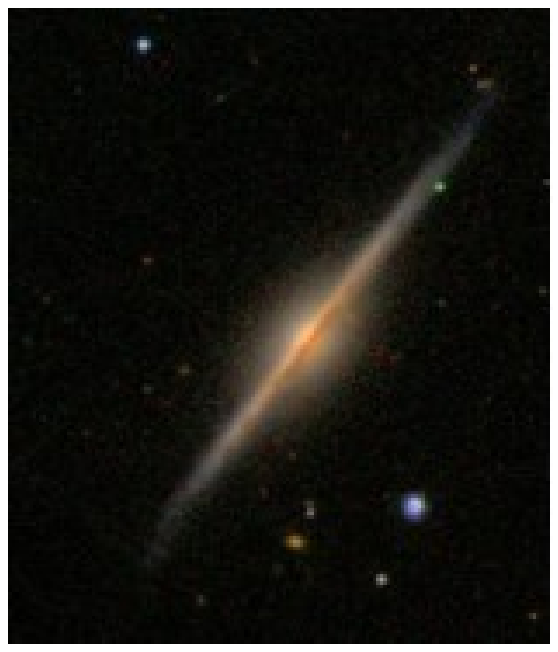}
\includegraphics[scale=0.85]{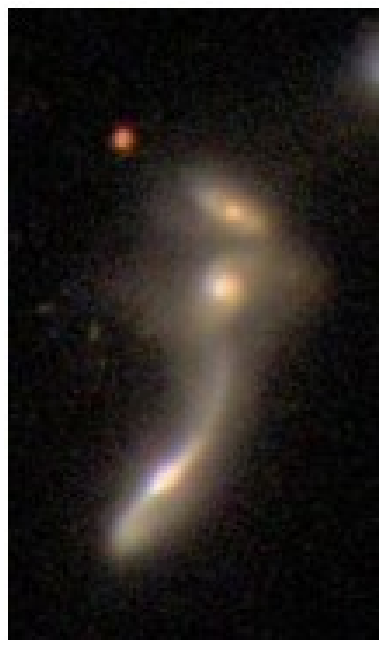}
\includegraphics[scale=0.85]{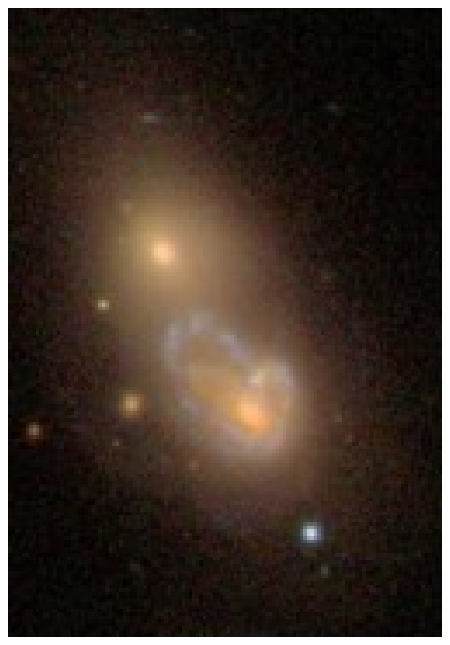}\\
\includegraphics[scale=0.85]{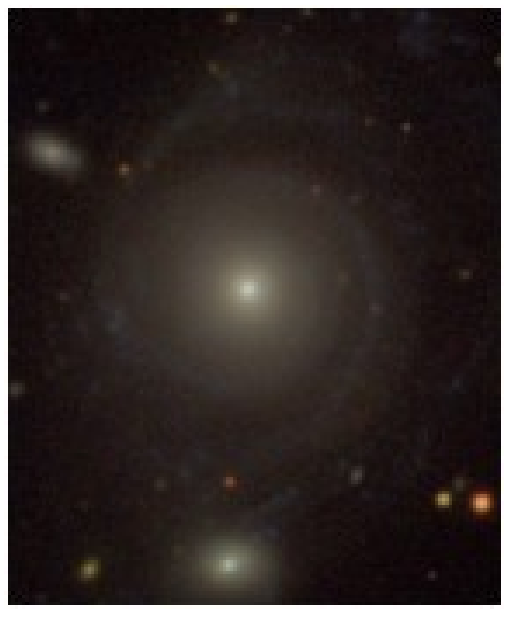}
\includegraphics[scale=0.85]{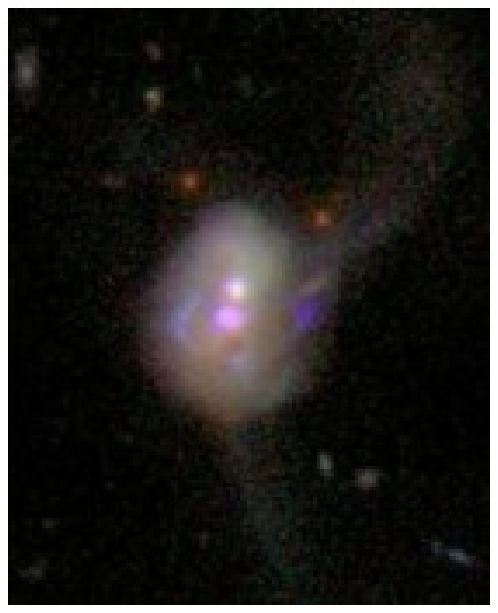}
\includegraphics[scale=0.85]{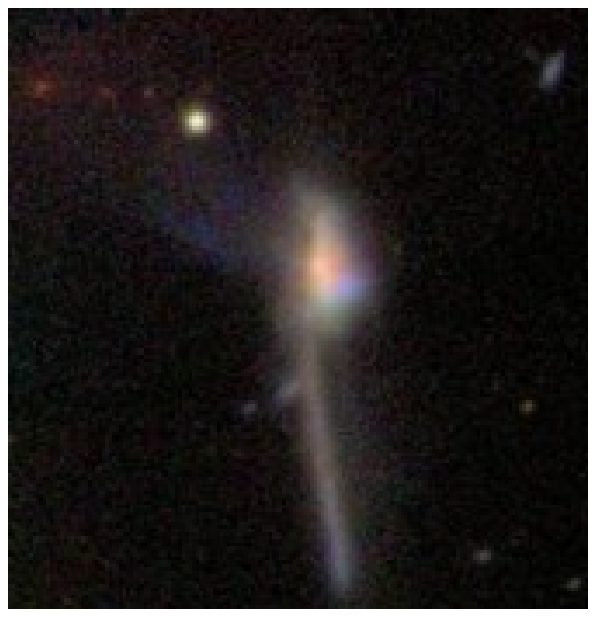}\\
\includegraphics[scale=0.85]{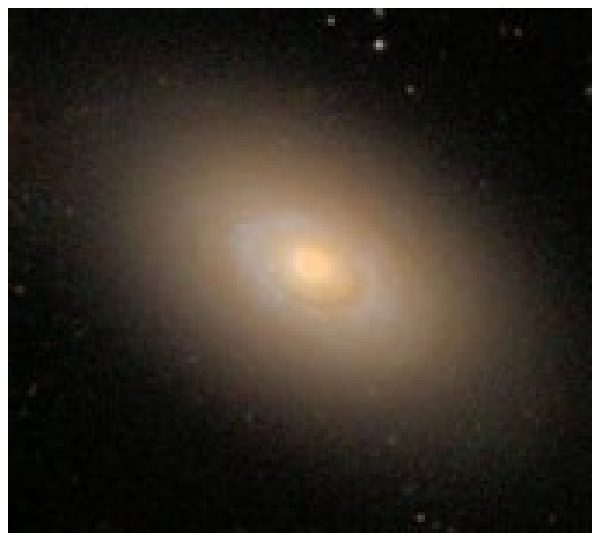}
\includegraphics[scale=0.85]{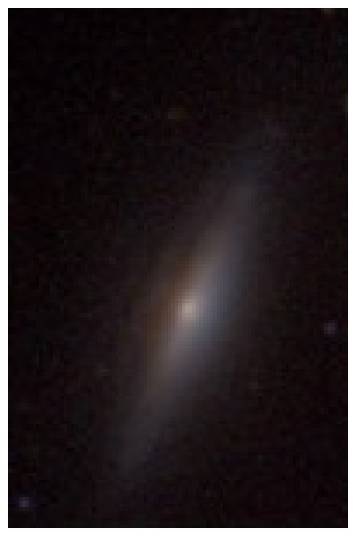}
\includegraphics[scale=0.85]{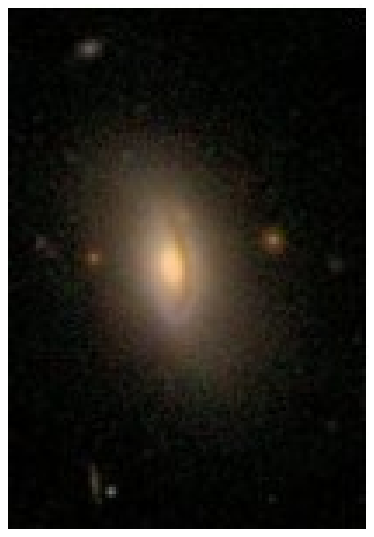}
\includegraphics[scale=0.85]{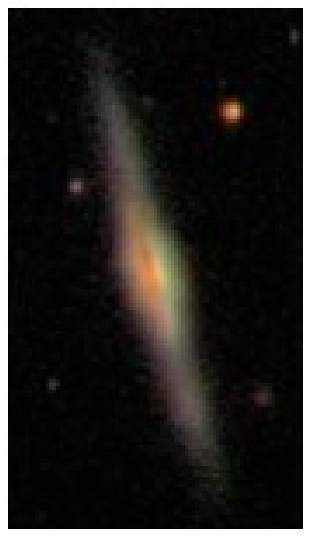}\\
\includegraphics[scale=0.85]{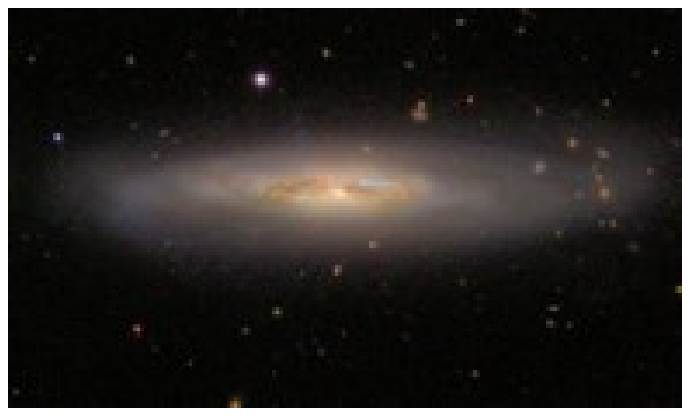}
\includegraphics[scale=0.85]{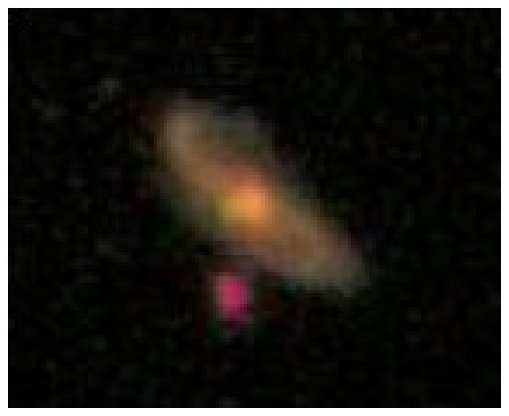}
\caption{$gri$ composite images of galaxies in the sample taken at
  Kitt Peak from M. Koss et al. (in preparation) and the SDSS. An arcsinh
  stretch was used as described in Lupton et al. 2004 with color
  scaled by flux.  Galaxies are from left to right$-$first row: IC 2461,
  Mrk 176, NGC 1142; second row: Mrk 348, Mrk 463, Mrk 273; third row:
  NGC 4138, ESO 506$-$G027, NGC 4074, MCG +08$-$21$-$065; fourth row: NGC
  4388, 2MASX J10335255+0044033.}
\label{figure:image}
\end{figure*}

Studying properties of AGN host galaxies is of particular interest to
understand a relationship between an AGN and its host. We made $gri$ composite
images of 12 galaxies in our sample, for which imaging data were taken at
Kitt Peak (M. Koss et al. in preparation) or the SDSS data are available. An arcsinh
stretch was used as described in Lupton et al. (2004) with color scaled by 
flux, and resulting images are shown in Figure \ref{figure:image}.

The morphology of the galaxies spans a wide range including early-type,
spiral, and strongly interacting galaxies. The inclination of the spiral
hosts also ranges from edge-on to near face-on.
Winter et al. (2009) studied host galaxies of  {\it Swift}/BAT-selected
AGNs and argued that AGNs with small absorption tend to
be near face-on, while highly absorbed AGNs are found regardless of the
inclination. The absorption columns for our sample are large ($N_{\rm H} =
10^{23-24}$ cm$^{-2}$), and the observed wide
variety of the inclination angles of the hosts is in agreement with the
results obtained by Winter et al. (2009). The three objects with the lowest values
of $f_{\rm scat}$ (IC 2461, ESO 506$-$G027, and MCG+08$-21-065$) are
edge-on spirals, while Mrk 348, which has similarly small $f_{\rm scat}$,
resides in nearly face-on spiral. Thus, the relation between
the inclination angle of the host and $f_{\rm scat}$ is not clear.

Four among 12 objects show clear signature of galaxy interaction
(Mrk 176, NGC 1142, Mrk 463, and Mrk 273). 2MASS~J10335255+0044033
is also possibly interacting. Three objects (Mrk 176, Mrk 463, and Mrk 273) 
show intermediate to large values of the scattering fractions, while 
$f_{\rm scat}$ for NGC 1142 is 0.8\%. The hosts of objects with
small $f_{\rm scat}$ ($<0.5$\%) are spirals without signature of
interaction (IC 2461, Mrk 348, ESO 506$-$G027, and MCG+08$-21-065$;
note that signatures of a double nucleus Mrk 348 were found in higher 
resolution $Hubble Space Telescope$ ($HST$) images by Gorjian (1995)).
Through interactions of galaxies, gas inside galaxies could be transported to
the central region. Such gas is a candidate for the source of obscuring
matter around AGNs. Our findings, however, suggest that geometrically thick
obscuring matter is not directly related to galaxy interaction.

Galaxy interaction also leads to active star formation in the galaxies,
and we expect intense soft X-rays from such starburst activity. Therefore,
we should note that $f_{\rm scat}$ could be overestimated in such systems.
Among the objects showing interaction, Mrk 463 and Mrk 273 are ULIRGs,
and significant starburst activity is likely to be taking place.
The comparison between $f_{\rm scat}$ and $f^{\rm corr}_{\rm scat}$ in
Section 4.1 shows that the soft X-ray emission in Mrk 273 is indeed
dominated by starburst. The differences between $f_{\rm scat}$ and
$f^{\rm corr}_{\rm scat}$ for the other objects (Mrk 176, NGC 1142, and Mrk 463)
are not very large, and the contribution of starburst to the soft X-rays does 
not alter the above discussion.

\section{Conclusions}
We derived a new sample of obscured AGNs from the $2XMM$ 
Catalogue paying attention to the strength of the
scattered emission. Our sample covers a wide range of scattering
fractions, which is a fraction of scattered emission with respect to
direct emission and reflects the opening angle of the obscuring torus, and
allows us to investigate relations between geometrical structure
around a nucleus (the opening angle of the torus) and multiwavelengh
properties of AGNs. To calculate the scattering fractions
quantitatively, we analyzed X-ray spectra obtained with {\it XMM-Newton} and
found that our sample covers the range $f_{\rm scat}$$\sim$0.1\%$-$10\%.
Optical spectra of five objects were also analyzed.

We investigated multiwavelength properties for our sample, and found
the followings.

\begin{itemize}
  \item There is no significant correlation between a scattering
    fraction and a far-infrared luminosity, though the absence of a
    correlation may be partly due to selection biases. This result
    implies that the energy generated by nuclear starburst activity is
    not a major source of energy supply to maintain the torus-like
    shape of obscuring matter around the nucleus. There is no
    significant correlation between an infrared color $f_{\rm
      60}$/$f_{\rm 25}$ and a scattering fraction.
  \item We found that the ratios of extinction-corrected
    [\ion{O}{3}]$\lambda$5007 to intrinsic 2$-$10 keV luminosities for
    objects with a small scattering fraction tend to be smaller than
    those with a large scattering fraction. This result is in
    accordance with our prediction that objects with a small opening
    angle (or a small scattering fraction) also have a small amount of
    narrow line region gas. Surveys using optical emission lines could
    be biased against buried AGNs.

 \item We compared black hole masses and Eddington ratios with
    scattering fractions. The comparison with black hole masses showed
    that there is only a hint of very weak correlation.  The Eddington
    ratio of buried AGNs tends to be larger for objects with a small
    scattering fraction. The scattering fraction could be a useful
    parameter to study growth and evolution of supermassive black
    holes.
    \item We examined optical images of 12 galaxies in our sample.
    No clear relationships between $f_{\rm scat}$ and the inclination
    angle or signatures of interaction of the hosts.
\end{itemize}

\acknowledgments We are grateful to Tohru Nagao and Yoshiaki Taniguchi
for useful discussions.  This paper is based on observations obtained
with {\it XMM-Newton}, an ESA science mission with instruments and
contributions directly funded by ESA Member States and the USA
(NASA). The Kitt Peak National Observatory images were obtained using
MD-TAC time for program 0417.  Kitt Peak National Observatory,
National Optical Astronomy Observatory, is operated by the Association
of Universities for Research in Astronomy (AURA) under cooperative
agreement with the National Science Foundation. This research made use
of the NASA/IPAC Extragalactic Database, which is operated by the Jet
Propulsion Laboratory, Calfornia Institute of Technology, under
contract with the National Aeronautics and Space Administration. This
work is supported by Grants-in-Aid for Scientific Research 20740109
(Y.T.), 20540230 (Y.U.), and 21244017 (H.A.) from the Ministry of
Education, Culture, Sports, Science, and Technology of Japan.

Facilities: \facility{{\it XMM-Newton}, {\it Swift}, {\it Sloan}, SAAO:1.9m, KPNO:2.1m}

\end{document}